# Grain boundary character distributions in nanocrystalline metals produced by different processing routes

David B. Bober[a,b], Amirhossein Khalajhedayati[c], Mukul Kumar[b], Timothy J. Rupert[a,c,]*

[a] Department of Mechanical and Aerospace Engineering, University of California, Irvine, CA 92697, USA

[b] Lawrence Livermore National Laboratory, Livermore, CA 94550, United States

[c] Department of Chemical Engineering and Materials Science, University of California, Irvine, CA 92697, USA

*E-mail: trupert@uci.edu

**Abstract**

Nanocrystalline materials are defined by their fine grain size, but details of the grain boundary character distribution should also be important. Grain boundary character distributions are reported for ball milled, sputter deposited, and electrodeposited Ni and Ni-based alloys, all with average grain sizes of ~20 nm, to study the influence of processing route. The two deposited materials had nearly identical grain boundary character distributions, both marked by a Σ3 length percentage of 23-25%. In contrast, the ball milled material had only 3% Σ3-type grain boundaries and a large fraction of low angle boundaries (16%), with the remainder being predominantly random high angle (73%). These grain boundary character measurements are connected to the physical events that control their respective processing routes. Consequences for material properties are also discussed with a focus on nanocrystalline corrosion. As a whole, the results presented here show that grain boundary character distribution, which has often been overlooked in nanocrystalline metals, can vary significantly and influence material properties in profound ways.

**Introduction**

High strength, wear resistance, and fatigue tolerance make nanocrystalline metals attractive structural materials.[1] For example, pure nanostructured Ti is being tested as a replacement for less biocompatible Ti-6Al-4V in medical implants.[2] Other current or near-term applications include more environmentally benign industrial hard coatings,[3] and alternatives to depleted uranium munitions.[4] To facilitate these advances, processing scientists have developed many techniques to synthesize nanocrystalline metals in a variety of forms and compositions. Bulk parts can be manufactured by the top-down refining of a coarse grained alloy into a nanocrystalline one by severe plastic deformation (SPD).[5] Severely deformed powders produced by ball milling can be consolidated into bulk forms, or bulk sections may be made directly through accumulative roll bonding (ARB), equal channel angular pressing (ECAP) or high pressure torsion.[6-8] A variety of physical, chemical, and electrochemical deposition techniques can be used to produce nanocrystalline coatings and even thin sheets.[9-11]

The current understanding of nanocrystalline metals has been primarily built around average grain size, $d$, driven by the past success of the Hall-Petch relation.[12] At fine grain sizes where the Hall-Petch relationship breaks down, it has been replaced by new scaling rules that again relate strength to grain size.[13] The transition from one scaling rule to another occurs at critical grain sizes where the dominant deformation mechanisms change. The first grain size threshold is 100 nm, below which dislocations nucleate at grain boundaries, sweeping through entire grains without interacting with each other and forming tangles.[14] At even smaller grain sizes, around 10 nm, grain boundary sliding and rotation supplant dislocations as carriers of plasticity, eventually leading to an inverse Hall-Petch slope.[14, 15] Similar grain size-based relationships have been applied to other mechanical properties, like wear and fatigue resistance, and functional properties, like magnetic coercivity and permeability.[16-18]

The common theme to the deformation mechanisms above is that grain boundary sites become increasingly important, yet characterization of nanostructured materials rarely focuses on the boundary itself. Expanding the characterization of nanocrystalline microstructures to include more grain boundary information may help address unanswered questions about structure-property relations and also open the door for control of such features in the future. Work on conventional, coarse-grained metals has demonstrated that grain boundary networks can control a wide range of properties, from fracture to corrosion.[19, 20] In ultrafine grained (UFG) metals, or those metals with grain size larger than 100 nm but smaller than 1000 nm, the fraction of high angle boundaries has been implicated as a possible key to enhancing ductility.[21] Such effects are expected to be exaggerated at nanoscale grain sizes, where a large fraction of atoms resides in the grain boundary region. Indeed, some of the only nanocrystalline work to explicitly consider grain boundary type has been investigations into the unexpected ductility of nanotwinned copper, which is now thought to be a consequence of those special boundaries' ability to provide both soft and hard directions for dislocation motion within a single grain.[22, 23] Barmak et al. have also shown that the choice of whether or not to consider twins as grain boundaries can strongly influence the calculation of grain boundary electron scattering coefficients in nanocrystalline copper.[24]

Since different nanocrystalline processing methods are controlled by a variety of physical mechanisms, there is reason to expect that these techniques will produce materials with different grain boundary networks. As a first example, consider the grain boundaries in a ball milled material, which are formed by using very large plastic strains to refine a coarse grained material into a nanocrystalline one.[25] When applied to FCC metals, large deformations subdivide grains though the accumulation of dislocations that form into low energy dislocation structures (LEDS).[26, 27] Continued deformation causes the misorientation across LEDS to increase, forming low angle boundaries, and eventually high angle ones.[28] Twin fragmentation is a complimentary mechanism which has been proposed for nanoscale refinement [29], where deformation twins form within existing grains and the narrow twins are then subdivided by LEDS. Several works on ball milling [30-32] cite another theory proposed by Hellstern et al. [33], who suggested that continued deformation drives nanocrystalline grain rotation, transforming low angle boundaries into high angle ones. The extent to which these competing mechanisms may operate is sure to affect the grain boundary network.

On the other hand, materials produced by deposition methods must be understood within a different framework. For both physical vapor deposition and electrodeposition, films form as atoms bond to the growth surface. This commonality causes similar structural development, even though one process is purely physical and the other is electrochemical. As new atoms deposit, they briefly undergo surface diffusion before being confined within the bulk. In nanocrystalline growth, adatoms are restricted to small rearrangements and cluster with only their immediate neighbors. Clusters grow outward until contacting adjacent grains, with grain size determined by the relative rates of nucleation and growth. As a result, the microstructure is determined by the kinetics of growth-surface phenomena.[34] In the absence of deformation or recrystallization, grain orientation is fixed at nucleation, i.e., before a grain has formed boundaries with most of its neighbors.[34] The implication is, that for non-epitaxial growth, random nucleation orientations will produce a random (Mackenzie) distribution of boundary misorientations.[35] For real films, a {111} fiber texture often develops to minimize the surface energy of atom clusters. In this case, the Mackenzie distribution can be modified to account for texture.[36] For columnar growth (granular-epitaxy), a misorientation distribution favoring low angle boundaries would be expected along the growth normal.

In this study, we present the grain boundary character distributions (GBCDs) of nanocrystalline Ni and Ni-based alloys produced by ball milling, sputter deposition, and electrodeposition. Our results show that these processes can produce very different boundary character distributions, with several samples of identical grain size used to highlight the structural diversity that exists independent of grain size. In addition, several process variables, such as ball milling duration, post-sputter annealing and electrodeposition reverse pulse current, are examined to help gauge the range of possible grain boundary character distributions for each method and explore what generalizations can be drawn. Grain boundary character distributions are assessed with transmission Kikuchi diffraction (TKD) and the results are connected to the structure–determining physical growth mechanisms associated with each material's processing method. The connections drawn between grain boundary character distribution and synthesis mechanism allow the results to have significance beyond the specific materials used. Along this theme, our results are used to explore previous studies on the corrosion of nanocrystalline Ni and Cu. We find that processing history, as a possible proxy variable for the grain boundary character distribution, can be a good predictor of nanocrystalline corrosion resistance.

**Methods**

Nanocrystalline Ni-based alloys were chosen for this study because they can be synthesized by multiple techniques. Samples were produced with ball milling, sputter deposition, and electrodeposition, all having an average grain size near ~20 nm. A constant grain size was targeted to help isolate the effect of processing on grain boundary character distribution. Ball milling was performed with a SPEX 8000M high energy mill, using a hardened steel vial and balls, and a ball to powder ratio of 10:1.  A process control agent of 1 weight % steric acid was added to moderate cold welding, which can otherwise cause excessive powder agglomeration.  Preliminary processing experiments showed that the material approached a steady state grain size after ~4 h of milling. Subsequent samples were milled for either 4 h or 10 h.  The milling operation was carried out under an Ar atmosphere to avoid O and N contamination.  Magnetron sputtering was used to deposit Ni films, 260±9 nm thick, onto Si wafers.  A 99.999% pure Ni target was used, which, coupled with vacuum processing, yields very pure nanocrystalline films.  A power of 350 W and argon pressure of 0.5 mTorr produced a deposition rate of 1.2 Å/s.  The substrates were rotated parallel to the target to maximize uniformity.  Deposition occurred near room temperature to obtain the desired grain size.  Pulsed electrodeposition of Ni-W was performed following the work of Detor and Schuh, which allowed a precisely tunable grain size.[37]  Increasing reverse pulse current density preferentially removes tungsten from the growth surface, lowering its concentration in the final film and increasing grain size.[37]  Reverse pulse currents of either 100 mA/cm$^2$ or 150 mA/cm$^2$ were used to deposit Ni-W films onto pure Cu substrates, with respective W contents of 12 at.% and 6.2 at.%.

A dual beam FEI Quanta3D scanning electron/focused ion beam (SEM/FIB) microscope was used for secondary electron imaging and to measure chemical composition using energy-dispersive X-ray spectroscopy (EDS).  SEM/FIB was also used to prepare TEM specimens using the in situ lift-out method.  Ball milled particles were embedded in epoxy and mechanically polished before FIB milling. Electrodeposited samples were FIB cut perpendicular to the growth direction, producing a cross section view of the microstructure.  Low angle Ar ion milling (Fischione 1010) at 2-3 kV and 5 mA was used to prepare plan view TEM foils from the sputtered samples, which were released from their silicon substrates with XeF$_2$ etching. All samples were briefly cleaned in a 10 W oxygen plasma prior to orientation mapping to mitigate carbon contamination (South Bay Technologies PC2000).

A long-standing impediment to more clearly understanding nanocrystalline grain boundary phenomena was the limited spatial resolution of electron backscatter diffraction (EBSD), which revolutionized the study of grain boundary networks in coarse grained alloys. Research into grain boundary populations was generally limited to grain sizes greater than 100 nm [38], although this boundary could be pushed using laborious transmission electron microscope (TEM) methods.[39] With the advent of precession electron diffraction (PED) and transmission Kikuchi diffraction (TKD), orientation mapping is now possible with ~2 nm resolution.[40, 41] Grain orientations were measured here with TKD, following Keller and Geiss [41], and Trimby.[42] The same SEM was operated at 30 kV and 11 nA, with diffraction patterns collected on an Oxford Instruments Nordlys F+ EBSD detector. The thin sample was held at a working distance of 3.5 mm and tilted 20˚ from horizontal, so that the upper side faced away from the EBSD detector. A 1 mm aperture was used, along with the microscopes high current analytical mode. Step sizes of 2-10 nm were used based on the grain size of the materials as individually specified in the Results and Discussion. Scan times were kept short to mitigate drift, generally around 5-10 minutes for the smallest grain sizes where it posed the largest problem. The smallest maps contained only 2,500 points, but a sufficient number of maps were collected that each specimen had a total of 64,000 to 178,000 points scanned. The total areas scanned for each material are individually specified in the discussion. In analyzing the TKD data, a minimum threshold of 2 degrees was selected for grain reconstruction. Each map was also processed with a standard dilation method that ensured unindexed pixels at grain boundaries were assigned an orientation. In doing this, a minimum grain size of 4 pixels was imposed to mitigate incorrectly indexed pixels. Examples of both the unprocessed maps and reconstructed maps will be presented to show the effect of this methodology. TKD measurements were supplemented by bright field TEM and selected area electron diffraction, which provided an overview of grain size, morphology and texture (Phillips/FEI CM20 and CM300).

The hardness of mounted and polished ball milled particles was measured with an Agilent G200 nanoindenter. The indents were performed at a constant strain rate of 0.05 s$^{-1}$ with a 10 s hold and a Berkovich tip. Shallow indentations (0.1 µm to 0.5 µm) were used to avoid the possibility of substrate effects, since the particles' average minor axis diameter was ~10 um. This depth range exhibited the well-known indentation size effect, with shallower indents yielding greater hardness.[43] A consistent 0.4 µm indentation depth was selected for ease of comparison, although the trends reported were also observed across the full range of depths.

**Results and Discussion**

To begin, we investigated the grain size of sputtered, electrodeposited and ball milled samples using bright field TEM. The ball milled sample that was milled for 10 h is discussed here. Presented in Figure 1(a-c), these TEM images show nanocrystalline grains of equiaxed shape and with a narrow size distribution. The average grain size appears similar across each image, which are shown at equal magnification. All three materials show strong diffraction contrast between neighboring grains. Selected area diffraction patterns were also collected, with the continuous diffraction rings in Figure 1(d-f) indicating a wide range of orientations are present in all samples, confirming that they each have many high angle boundaries. Together with the equiaxed grains and strong diffraction contrast, this diffraction pattern implies that the ball milled material is not a cellular or low angle structure, as is sometimes observed in less deformed ECAP specimens.[6] In the ball milled material, weak diffraction contrast within grains is due to small misorientations caused by the extensive deformation history. The W atoms dissolved in the electrodeposited sample swells the Ni lattice, which is manifested in the smaller ring diameters in Figure 1(e).[37] Overall, the TEM images and diffraction patterns show relatively similar microstructures, although a few key differences in composition and deformation history are also apparent. The similarity is reinforced by quantitative analysis of the grain size using TKD. It was found that the ball milled, electrodeposited, and sputtered samples had mean grain sizes of 23 nm, 22 nm and 20 nm, respectively. All three materials have a narrow grain size distribution, which is shown by the cumulative distribution functions in Figure 2.

To measure how grain boundary character distribution depends on the processing technique used, orientation maps were collected with TKD. Examples of these maps for each material are shown in Figure 3(a-c), with no post-processing having been performed. Each map was collected with a 2 nm step size, with a total of 142,000, 64,000 and 75,000 points being scanned for the ball milled, electrodeposited and sputtered materials, respectively. The colors in Figure 3(a-c) correspond to different orientations, as labeled on the inverse pole figure legend. The black areas are points that could not be indexed, typically because the probe spanned a grain boundary and two Kikuchi patterns were convolved. Coincident site lattice (CSL) theory, which groups grain boundaries according their ideal number of shared lattice sites, was used to categorize boundaries by applying the Brandon criterion.[44, 45] The low angle ($\Sigma1$), twin ($\Sigma3$), twin-related ($\Sigma9$ and 27) and other low-CSL ($\Sigma1$-29) grain boundary fractions were chosen for inclusion in the GBCD because they are the easiest to relate to processing history and material properties. The post-processed TKD maps showing reconstructed grain boundaries are shown in Figure 3(d-f). The areas of these maps match those presented above in Figure 3(a-c). Blue lines represent low angle boundaries (2-15°), while red is used for Σ3s and black for all others (>15°). The gray scale contrast corresponds to the Kikuchi pattern quality.

The GBCDs for each material were calculated from the boundary length fraction in the TKD maps. They are shown in Figure 4, whose most striking feature is how similar the sputtered and electrodeposited GBCDs are, while the ball milled GBCD is noticeably different. For example, the ball milled sample has a much larger percentage of Σ1 boundaries (16%) than the two deposited materials (7-8%). In contrast, the two deposited samples have much higher Σ3 percentages (24-25%), as compared to the ball milled material (3%). This Σ3 fraction reflects twin-type boundaries of both coherent and incoherent varieties. The higher Σ9,27 and Σ1-29 boundary fractions in the deposited materials are a direct consequence of the high Σ3 fraction, because Σ3 interactions produce more $\Sigma 3^n$ boundaries and boost the overall Σ1-29 fraction.[46] These observations can be understood by focusing on each processing route in turn.

*Ball Milled Ni*

The fraction of Σ1 boundaries in the ball milled material is notably large compared to the materials produced by deposition. To understand this observation, it is useful to consider ultrafine grained (UFG) materials produced by ECAP and accumulative roll bonding. Like ball milling, these methods use severe plastic deformation to refine the grain size, although they cannot achieve grain sizes (*d*) below about 100 nm.[8, 47] For ECAP materials, the Σ1 percentage typically reaches a minimum plateau of 20-40% for plastic strains greater than about 6.[39, 48, 49] A similar result has been found for accumulative roll bonded Ni, where the generation of high angle boundaries slows considerably as strains reach about 6, although no true plateau was reached in the work of Zhang et al.[8] From this perspective, the 16% Σ1 percentage found after 10 h of ball milling is only slightly low. The fact that the $\Sigma 1$ content at $d \sim 20$ nm is comparable to that commonly measured at $d > 100$ nm suggests that no new mechanism for creating random high angle grain boundaries (RHAGB) becomes operative during nanocrystalline grain refinement (100 to 20 nm). This opposes the notion that grain rotation provides an added mechanism to increase RHAGB fraction during the severe plastic deformation of nanocrystalline materials.[33] Instead, it supports the continuity of grain refinement mechanisms crossing the UFG and nanocrystalline scales, as proposed by Hughes and Hansen.[50] Hughes and Hansen looked at the boundary network morphology caused by sliding contact and observed universal scaling features which suggested a continuity of subdivision mechanism until at least 10 nm. Similarly, the ball milled material's $\Sigma 3$ percentage (3%) is also in the range reported for UFG Ni produced by SPD (3-5%).[48, 49] Consequently, the total percentage of random high angle grain boundaries (RHAGB) in the ball milled material was 73%, which is in the (upper) range observed for SPD UFG metals.[39, 48, 49] This is consistent with more qualitative selected area electron diffraction measurements in other ball milled metals, where continuous selected area diffraction rings indicate a random high angle grain boundary structure.[31] Nanocrystalline grain refinement by the LEDS mechanisms would explain why the ball milled GBCD is so similar to that observed in UFG nickel, where grain subdivision by LEDS is well established.[27]

The orientation distribution can also help explain how the ball milled GBCD developed. Each dot within the stereographic projection in Figure 5(e) (bottom right) corresponds to the orientation of a pixel from the TKD scan. Their inhomogeneous distribution indicates that some orientations are more likely than others. Without even considering the specific texture, the presence of preferred orientations in Figure 5(e) already gives an indication of how the deformation proceeded. The non-random texture suggests that dislocation plasticity was more significant than grain boundary sliding or grain rotation, because the former mechanism promotes texture development while the latter two disrupt it. However, the exact balance between them is difficult to determine, although estimates for simple deformation geometries have been made using synchrotron data, forward texture modeling, and analysis of grain shape.[51] The LEDS and twin fragmentation mechanisms are both based on dislocation plasticity and are consistent with finding a preferred texture. Examining the specific texture for more insight is complicated by the nature of ball milling, and to a lesser extent, sample preparation. In Figure 5(e), the {110} fiber is roughly oriented along the 'y' direction, but unfortunately the FIB extraction of TEM foils from polished powder particles leaves this direction without a physical meaning. Ball milling does not have a well-defined deformation geometry, so relating a specimen axis to strain history is impossible.

A second ball milled sample was created to provide a snapshot of the microstructure at an earlier stage of its evolution. This second sample was only milled the 4 h needed to approach the minimum grain size. Bright field TEM images after milling 4 h and 10 h look quite similar, as shown in Figures 5(a) and (c), respectively. Examples of the TKD maps for this material are shown in Figures 6(a) and (b). They were collected at a step size of 3 nm, with a total of 178,000 points scanned. TKD measurements confirmed that the mean grain size only decreased by a small amount during the additional 6 h of milling, from 27 nm at 4 h to 23 nm at 10 h. Comparing the GBCD produced at short (4 h) and long (10 h) milling times reveals several differences which are plotted in Figure 5(b) (top center), where the length percentages of CSL categorized boundaries are plotted. At longer milling times, the $\Sigma 3$ fraction is lower than at short times. This finding is in keeping with grain refinement by dislocation subdivision or LEDS, which promotes a gradual increase in boundary misorientation and does not especially favor the development of $\Sigma 3$ boundaries.[28] On the other hand, the $\Sigma 1$ fraction increased slightly with milling, which is unexpected as other SPD materials often experience a reduction in the Σ1 fraction as dislocations accumulate to produce RHAGBs.[39]

The effect of milling time is also evident from the powder morphology, as shown in the SEM micrographs in Figure 7. These secondary electron images show the size, shape and surface texture of the individual powder particles. After 4 h of milling the particles had a platelet-like morphology (Figure 7(a)), while after 10 h each particle appears to be composed of many small fragments cold welded together (Figure 7(b)). Note that Figure 7(a) appears at a slightly higher magnification than Figure 7(b). Also observed by Xun et al. [52], this change in particle shape is thought to be due to a shift in the balance between cold welding and fracture. The new morphology implies a new deformation geometry for subsequent impact events and is probably responsible for the development of different textures after 4 h and 10 h of milling, see Figures 5(d) and (e). Again, it is impossible to relate the texture to a well-defined deformation axis.

Two other notable changes that occurred with increasing milling time were an increase in Fe contamination from the hardened steel milling media and an increase in hardness. The Fe content rose from 5 at.% at 4 h to 18 at.% after 10 h milling, as measured with EDS. This type of contamination is common in ball milling, with similar high concentrations reported elsewhere.[53] Accompanying this change in composition, hardness also rose from 8.2 GPa to 9.1 GPa. Applying the solid solution strengthening model that Rupert et al. [54] have proposed for nanocrystalline metals, the change in Fe concentration would be expected to increase hardness by ~700 MPa. Grain boundary strengthening can also be expected to contribute, since the average grain size fell slightly during extended milling.[55, 56] Varied results have been reported on the relationship between hardness and grain size for nanocrystalline Ni.[57-59] Interpolation within each of these reports would predict an increase in hardness ranging from 140 to 1900 MPa, for a change from 27 to 23 nm grain size.[57-59]  Summing the solid solution and grain boundary strengthening effects leads to an expected increase in hardness of 840 to 2600 MPa.  One possible reason why the observed hardening (900 MPa) is at the lower end of the expected range is that not all grain boundaries contribute equally to grain size strengthening; low angle boundaries are less effective strengtheners.[60]  The studies interpolated to predict grain size strengthening [57-59] are based on deposited materials, which our GBCD results indicate tend to have more high angle boundaries than the ball milled materials.  This could lead to an overestimate for the predicted grain size strengthening effect in the ball milled sample because its grain boundary network has less strong obstacles to plasticity.

However, the ball milled material's RHAGB fraction is still quite high.[39, 48, 49]  Based on UFG studies, this will promote ductility because RHAGB's are more susceptible to grain boundary sliding.[21, 61]  Molecular dynamics simulations indicate that this enhancement might be even greater in nanocrystalline materials, given the increased grain boundary area.[62]  However, the high RHAGB fraction is also likely to leave the microstructure more vulnerable to coarsening because RHAGBs are more prone to thermal migration than are low energy boundaries.  This concern has particular relevance to ball milled powders, which must be consolidated at high temperatures if fully-dense bulk materials are desired.

*Electrodeposited Ni-W and Sputter Deposited Ni*

Having tried to place the ball milled GBCD in the context of SPD and literature on comparable materials, the same will be attempted for the two deposited materials.  First, it is instructive to recall the similarities in GBCD between the sputtered and electrodeposited samples, which both had high Σ3 fractions.  For a different perspective on the types of grain boundaries present, the frequency of their misorientation angles can also be plotted, as shown in Figure 8. Because deposited grains should have no knowledge of their in-plane neighbors until they coalesce, at which point their orientations are fixed, the grain boundary misorientation distribution (correlated) should be equivalent to the misorientation distribution for any set of randomly selected orientation pairs (uncorrelated). In contrast to this ideal case, the real correlated misorientation distribution deviates strongly from the randomly selected, uncorrelated, one around a misorientation of 60 degrees. This implies that a growth mechanism must exist that promotes this particular misorientation, which matches the $\Sigma 3$ type boundary. In fact, similar results are almost universally observed for other FCC materials that exhibit annealing or growth twins. As atoms deposit on {111} planes, some fraction occupy higher energy sites corresponding to stacking faults.[63, 64] Limited diffusion and high deposition rate can act to trap them in these locations, with subsequent layers growing the twin thickness.[63, 64] While these twins are low energy boundaries, they result from the incorporation of high energy planar defects and are intimately tied to the same non-equilibrium processes that create the nanocrystalline structure. Such an explanation accounts for both the high $\Sigma 3$ fraction and the misorientation peak at 60 degrees.

The high fractions of $\Sigma 9$ and $\Sigma 27$ boundaries result from the crystallographic network constraints imposed by the many $\Sigma 3$ boundaries.[46] This stems from the triple junction boundary combination rule, which requires the boundary misorientations meeting at a junction sum to zero. In other words, proceeding through each rotation once must result in the starting orientation. As more $\Sigma 3^n$ boundaries enter the microstructure, they are more likely to terminate at other $\Sigma 3^n$s, which requires the third boundary at the junction to also be a $\Sigma 3^n$.[46] This analysis rationalizes the similar GBCD of the electrodeposited and sputtered films in terms of shared growth mechanisms, which might be somewhat surprising because of their different compositions. Given that W addition lowers the stacking fault energy of Ni [65], it would have been reasonable to expect more growth twins in the Ni-W. The fact that this is not the case emphasizes the importance of kinetics in driving structural development during nanocrystalline deposition.[64]

The high excess energy of grain boundaries and their high volume fraction in nanocrystalline materials provides a powerful driver for grain growth. The situation is only exacerbated by the often non-equilibrium character of nanocrystalline grain boundaries, with notable examples including high energy interface planes and high grain boundary dislocation densities.[6, 66, 67] In the case of electrodeposited copper, room temperature grain coarsening has even been observed.[68] Despite being so common, it is unclear how grain growth affects nanocrystalline grain boundary character distribution. To investigate this question, a sputtered Ni sample was annealed at 524 K (250 °C) for 1 h. This temperature was selected because past experience with this material showed that annealing above about 374 K (100 °C) would cause a few grains to grow rapidly, consuming the surrounding nanocrystalline ones. Examples of the TKD maps collected for this material are shown in Figures 9 (a) and (b). They were collected at a step size of 10 nm, with a total of 90,000 points scanned. Figure 10 presents TEM images and TKD data from as-deposited and annealed sputtered Ni films. Many classical annealing twins are observed inside the larger grains which were not present in the original materials, as shown in Figures 10 (a) and (c). They can be recognized as the narrow bands within grains having straight edges along {111} planes and alternating contrast. Their presence caused the annealed grain boundary network to show a much higher twin length percentage (37%) and consequently higher total special percentage, as shown in Figure 10(b). The overall effect is a GBCD indicative of lower energy density. These changes can be explained with conventional theories of annealing twin formation. Annealing twins are believed to form from nucleation accidents occurring on the

{111} oriented steps of migrating grain boundaries, which subsequently propagate into the parent grain.[69, 70] Gleiter's model for annealing twins predicts the observed increase in twinning with increasing grain size.[70] Liu et al. [71] recently checked this model across grain sizes spanning the nanocrystalline and UFG regimes, finding good agreement. Gleiter's model also predicts significant twinning for both Cu and Ni, despite their different stacking fault energies.[70] The high number of twins may have also been enabled by the low temperature at which grain growth occurred. At these temperatures, growth accidents are less likely to be eliminated by diffusion.[69] The necessary boundary migration is still possible because of the strong driving force provided by the excess nanocrystalline grain boundary energy. This type of GBCD modification through annealing twins is also consistent with the unchanged fiber texture evident in Figures 10(d) and (e).

To further explore nanocrystalline deposition and the range of possible grain boundary character distribution, a second electrodeposited material was created with a larger grain size of 117 nm. Examples of the TKD maps for this material are shown in Figures 11(a) and (b). They were collected at a step size of 5 nm, with a total of 151,000 points scanned. The random texture remained unchanged across both samples (Figures 12(d) and (e)). The larger grained material contains more growth twins, as shown in the TEM images in Figure 12(a,c). They are again recognizable as narrow bands with alternating contrast that traverse grains. The effect of these twins is to dramatically increase the $\Sigma 3$ fraction to 51%, roughly double that of the 20 nm sample (25%), as shown in Figure 12(b). This change in $\Sigma 3$ fraction has a surprising dependence of W content, with more twins found at lower W content. The addition of W lowers the alloy's stacking fault energy [65], which would be expected to promote twinning. That the opposite is observed suggests that the origin of these $\Sigma 3$'s is likely to be related to the growth kinetics. The increase in Σ3 fraction with deposited grain size is somewhat analogous to the behavior observed under annealing, although in this case the boundary between a growing and shrinking grain is replaced by the growth front. With increasing reverse pulse current, the rate of twin nucleation on the growth front remained high, even while the rate of grain nucleation dropped.

*GBCD and Corrosion*

The high special fraction (67.7%) observed in the 117 nm grain size electrodeposited sample should have consequences for material properties. For example, a material with this special fraction is likely to inhibit intergranular corrosion.[72] This implies that the corrosion resistance of the 117 nm grain size sample is likely to surpass that of the ~20 nm grain size sample, if only GBCD is considered. Similarly, the 117 nm sample is also likely to be more corrosion resistant than coarse gained Ni, which tends to have a low special boundary fraction, unless processed in specialized ways.[19] In fact, prior literature shows just such a result; nanocrystalline Ni can exceed the corrosion resistance of coarse grained Ni.[73] Without considering boundary character, this may have seemed counterintuitive because nanocrystalline Ni has a high density of grain boundaries, which are generally more susceptible to corrosion than grain interiors.[74] This apparent anomaly led Roy et al. [38] to investigate the GBCD of nanocrystalline Ni with EBSD, where they were able to create orientation maps that captured only the larger grains present in the microstructure and saw increased twinning. From this, Roy et al. [38] concluded that a high special fraction was likely responsible for nanocrystalline Ni's corrosion resistance, a conclusion which is bolstered by the higher resolution data presented here. A similar finding was also recently reported by Zhao et al. [75] for nanotwinned copper, in which higher levels of twinning promoted corrosion resistance. In a columnar microstructure, it was found that having many twins parallel to the growth face reduced corrosion, although changes in {111} texture strength were also believed to contribute.[75]

While our data supports the conclusions of Roy et al. [38], it also points to the fact that high special boundary fractions are not universal in nanocrystalline metals and therefore increased corrosion resistance should not always be expected. In fact, even among studies on nanocrystalline Ni, conflicting grain size-corrosion relationships have been observed in both NaCl and $H_2SO_4$ electrolytes.[76-80] These results have been plotted in Figure 13, with the corrosion currents for each study normalized by the minimum corrosion current reported, to allow for easy comparison. Note that the grain size is plotted logarithmically to accommodate the large range. Examining the trends in Figure 13, it is clear that even for a single electrolyte, no consistent correlation with grain size exists. This could be due to any number of additional variables, including composition, texture, residual stress, and, of course, grain boundary network. These confounding factors make it difficult to determine the role of the grain boundary network, but Figure 13 motivates the search for a case where this might be more tractable. Data on the corrosion of nanocrystalline Ni produced by SPD would provide a helpful point of comparison because of the different GBCD expected, but this information is unavailable.

For an example where the role of grain boundaries and processing route is clearer, it is instructive to consider the case of nanocrystalline Cu. While so far we have focused on Ni and its alloys, Cu is another FCC metal in which the same microstructural formation mechanisms, such as grain refinement by LEDS and twinning by growth accidents, operate. The similarity between mechanisms provides some confidence that nanocrystalline Cu will show similar trends in GBCD as were observed in Ni. To recap, that would mean that severely deformed nanocrystalline copper is likely to have a preponderance of RHAGBs, while the deposited materials should be characterized by a high special fraction. This is turn should manifest as a difference in corrosion behavior, namely that deposited nanocrystalline Cu should be more corrosion resistant than Cu prepared by SPD. Fortunately, many corrosion studies of Cu exist that let this hypothesis be explored. For a uniform comparison, only those studies which used neutral or acidic chloride electrolytes (NaCl, HCl, Hanks solution) will be compared, because the corrosion of a metal is highly dependent on the corrosive media.[75, 81-85] Also to aid comparison, the nanocrystalline corrosion currents reported have been rescaled as the fractional difference from the coarse grained corrosion currents reported in each work. The intent is to minimize the effect of experimental differences, such as electrolyte concentration. The resulting relative corrosion rates are plotted in Figure 14 with data grouped by process type, either deposition or severe plastic deformation.[75, 81-85] It is obvious that the deposited materials are, as a group, more corrosion resistant that the plastically deformed ones.

This being a cross-study correlation, it is prudent to examine all factors that may be responsible for the observed behavior. In Figure 14, each marker has had its area scaled to be proportional to grain size, per the legend along the right side. From this, it can be seen that the trend holds across a range of grain sizes and that the process grouping is not merely serving as a proxy variable for grain size. Differences in impurity content could also be playing a role, as their segregation to grain boundaries can significantly enhance corrosion. Palumbo et al. [86] have attributed the (sometimes) superior corrosion performance of nanocrystalline metals to the dilution of impurities across a greater grain boundary area. However, electrodeposited materials in general have a higher impurity level than those prepared by equal channel angular pressing, which was the method of choice for every SPD study in Figure 14. If impurities were the controlling factor, a higher corrosion rate would be expected for the deposited group, but the opposite trend is found in

Figure 14. With regards to strain energy, it could be hypothesized that the higher strain energy of the severely deformed metals would promote corrosion, as it does for coarse-grained metals. However, Valiev et al. [87] have shown that short recovery heat treatments, which relax grain boundary structure and reduce strain energy, do not markedly improve the corrosion resistance of SPD materials. Differences in texture could also contribute to the corrosion resistance of the deposited metals. In the case of Inconel 600, Schuh et al. [88] showed that the corrosion rate of individual grains was proportional to their deviation from a {111} surface normal. One could therefore expect deposited materials, which often exhibit a {111} fiber texture, to exhibit enhanced corrosion resistance. Unfortunately, this is not an effect which is easy to control for, with Zhao et al. [75] showing how grain boundary character and texture have closely coupled effects on corrosion. In their work, nanotwinned copper was found to be more resistant to corrosion, with the effect being proportional to the extent of twinning.[75] However, the strength of the {111} texture was also proportional to the extent of twinning and distinguishing the effects proved challenging.[75] Most of the studies analyzed for Figure 14 do not include any texture information, which leaves open the possibility that the trend is due to a surface effect, rather than a grain boundary one.[75, 81-85] However, that conclusion presupposes that all of the deposited materials in Figure 14 have a {111} texture, even though there are many probable alternatives.[89]

An interesting case that supports attributing the trend in Figure 14 to the GBCD is that of surface mechanical attrition of stainless steel.[90] It has been reported that a highly twinned nanocrystalline region of the sample was more corrosion resistant than the bulk, while a similarly grain sized area with a random boundary network was less resistant than the bulk.[90] While additional experiments would help bolster these conclusions, coupling our data to prior literature suggests that synthesis method affects the grain boundary networks of nanocrystalline metals and that these features can help explain some of the variation in corrosion resistance that has been observed in past studies. This is in addition to the many other predicted effects of grain boundary type at the nanoscale, which notably include ductility and thermal stability.[21, 66]

**Conclusions**

In summary, a systematic study of the grain boundary character distributions of several nanocrystalline metals, produced by different processing routes but with similar mean grain sizes, has been presented. It was observed that electrodeposition and sputtering produced similar GBCDs, due to their shared structure-determining kinetics, with growth accidents on {111} planes responsible for the high $\Sigma 3$ fraction [26, 27, 63, 64] that in turn leads to a high $\Sigma 9,27$ fraction.[46] These similarities were found in spite of differences in composition which should affect the stacking fault energies of the materials. The ball milled material exhibited a GBCD that was much different than the deposited GBCDs, with very few twins and many more $\Sigma 1$ boundaries. The GBCD of the ball milled material was quite consistent with prior reports on UFG metals, suggesting a continuity of the LEDS grain refinement mechanism to nanoscale grain sizes.[26, 27, 39, 48, 49, 63, 64] The ball milled GBCD also varied somewhat with milling time, even as grain size stayed roughly constant. These results were then applied to consider some prior literature on nanocrystalline corrosion, finding that differences in GBCD may explain why deformed and deposited nanocrystalline materials exhibit different corrosion resistance. The connection of GBCD to processing method and important material properties will hopefully motivate additional future work on other nanostructured materials. Such experiments will become more commonplace as TKD and PED become increasingly available to researchers. As illustrated by the case of nanocrystalline copper corrosion, the GBCD can have a predictive power that can be even more important than grain size.

## Acknowledgements

We gratefully acknowledge support from the National Science Foundation through a CAREER Award No. DMR-1255305. This work was partly performed under the auspices of the U. S. Department of Energy by Lawrence Livermore National Laboratory under Contract DE-AC52-07NA27344. D.B.B. and M.K. were supported by the U.S. Department of Energy (DOE), Office of Basic Energy Sciences, Division of Materials Science and Engineering under FWP# SCW0939. D.B.B. also acknowledges the support of the Livermore Graduate Scholar Program at Lawrence Livermore National Laboratory during part of this work.

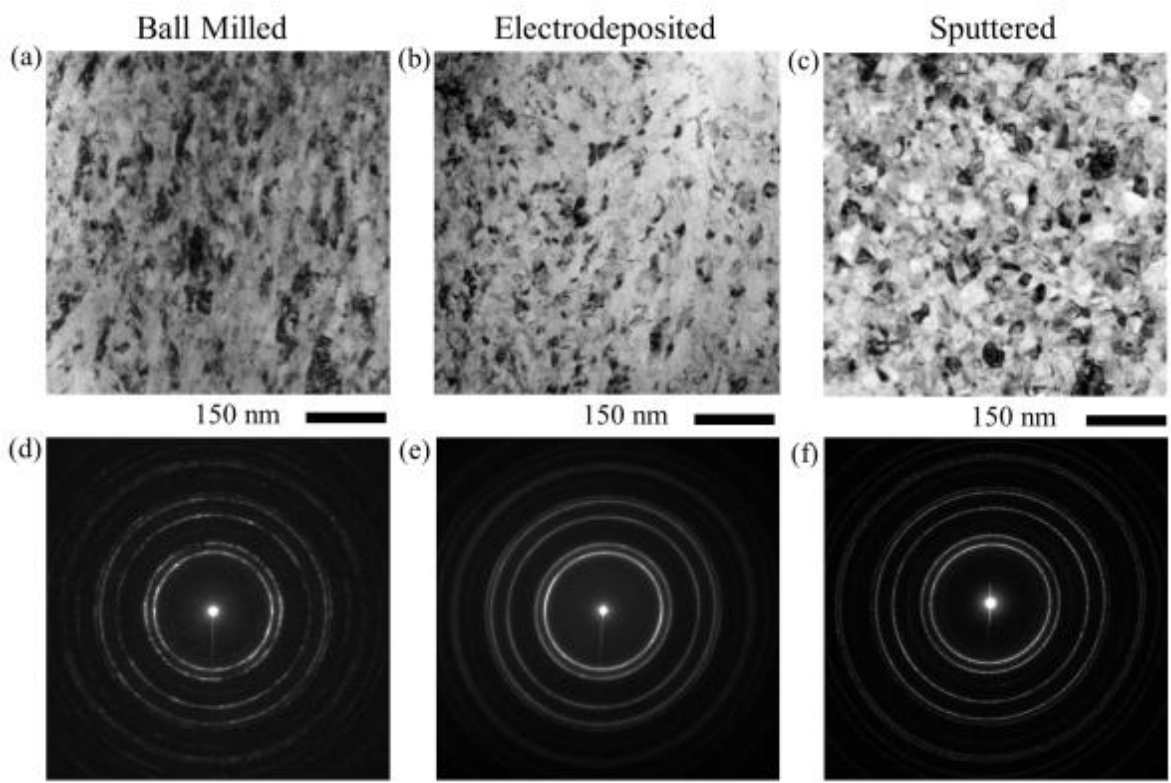

**Figure 1: Bright field TEM images of (a) ball milled, (b) electrodeposited, and (c) sputtered Ni and Ni alloys, with accompanying selected area diffraction patterns shown below in (d-f). Overall, the three materials have many similar characteristics, such as equiaxed grains averaging ~20nm.**

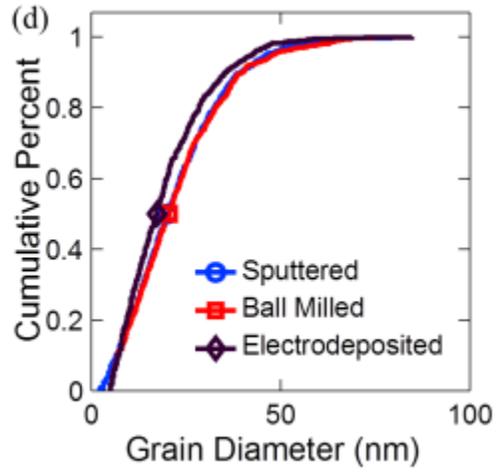

**Figure 2: The cumulative grain size distributions for each material confirm their similar grain sizes and distributions, all having mean grain sizes in the range of 20-23 nm.**

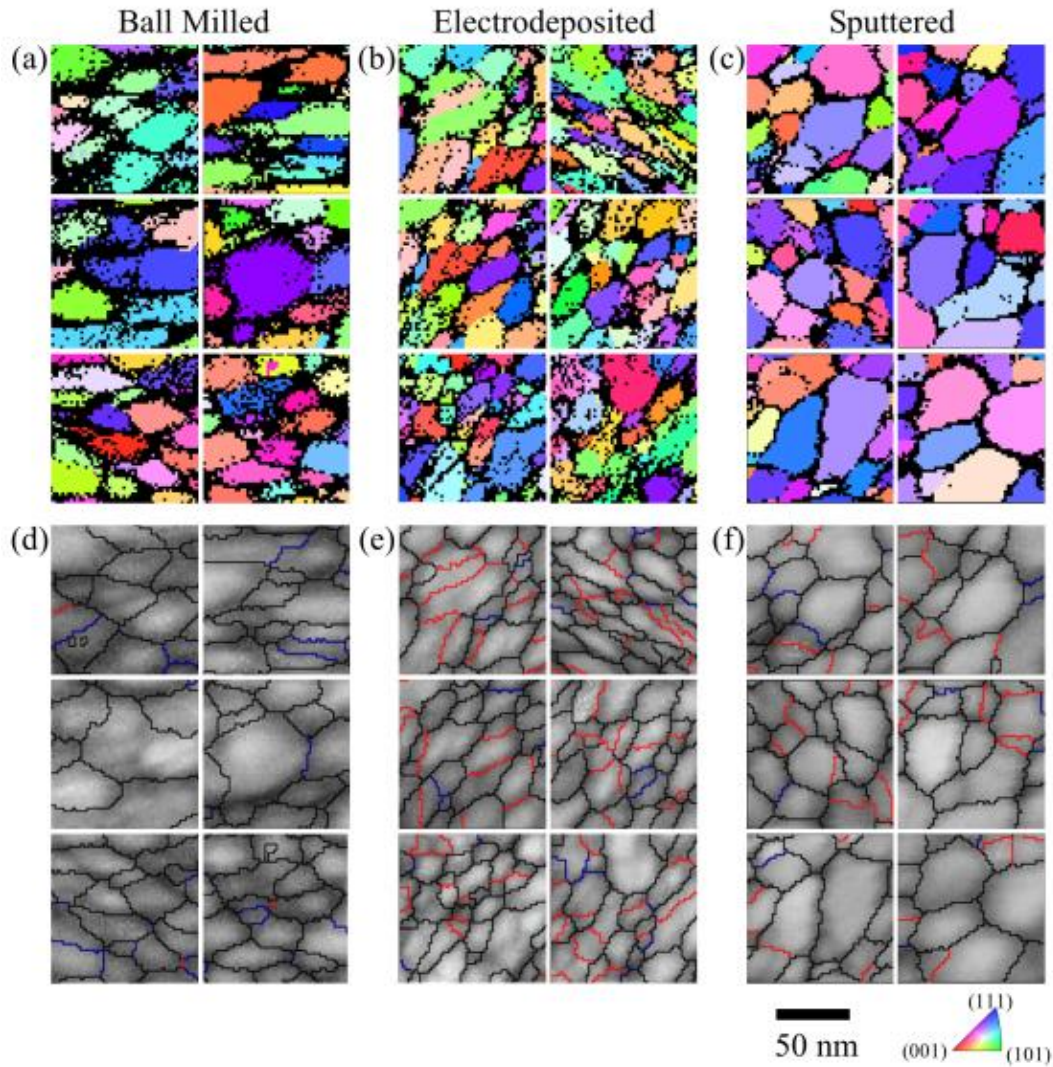

**Figure 3: Grain orientation maps for the (a) ball milled, (b) electrodeposited, and (c) sputtered materials. Reconstructed grain boundaries are shown for the same areas in (d-f). Red lines are Σ3 boundaries, blue are Σ1s and all other boundaries are shown in black. The grey scale contrast is the Kikuchi pattern quality.**

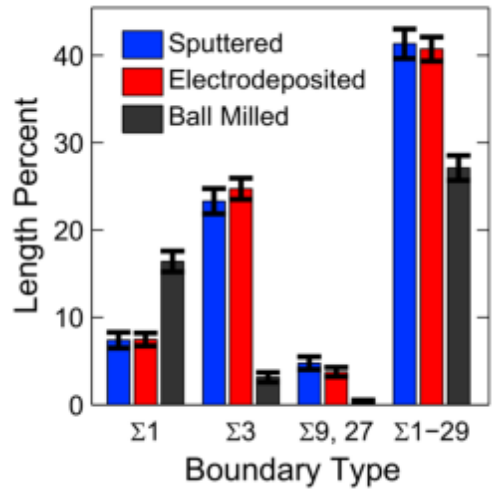

**Figure 4: Grain boundary character distributions for sputtered, electrodeposited, and ball milled samples with a constant grain size. The length percentages of several CSL-type boundaries are shown.**

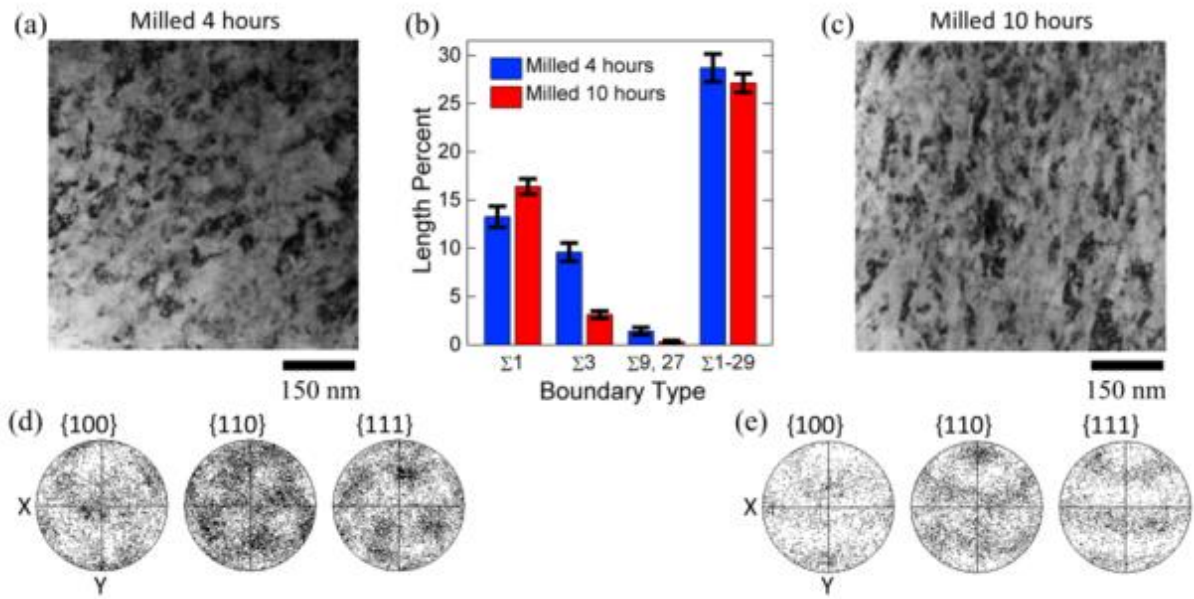

**Figure 5: Bright field TEM images and TKD pole figures for samples ball milled for (a, d) 4 h and (c, e) 10 h, along with their GBCDs (b).**

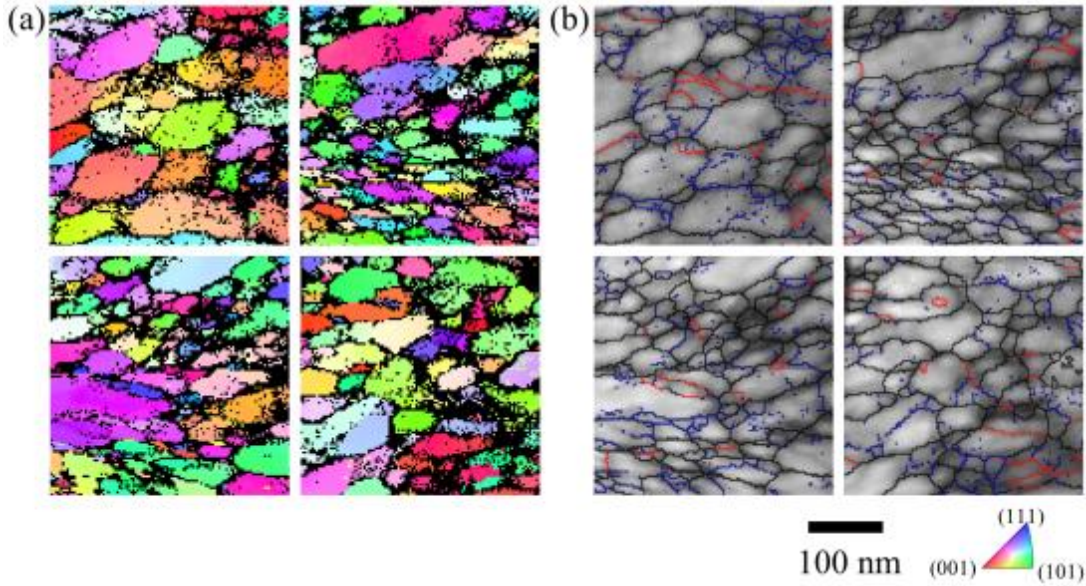

**Figure 6: (a) Grain orientation maps and (b) reconstructed grain boundaries for the 4 h ball milled material. Red lines are Σ3 boundaries, blue are Σ1s and all other boundaries are shown in black. The grey scale contrast is the Kikuchi pattern quality.**

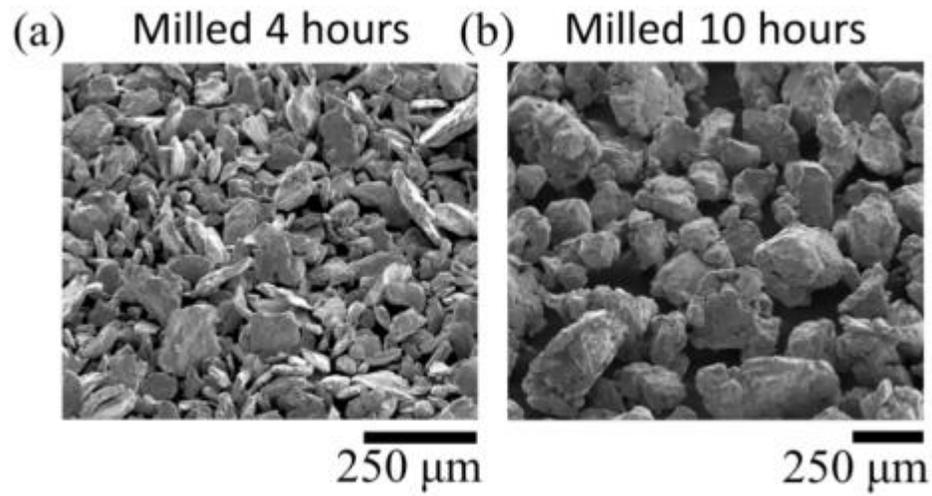

**Figure 7: SEM images showing a change from (a) platelet shaped particles after 4 h of milling to (b) larger equiaxed particles after 10 h.**

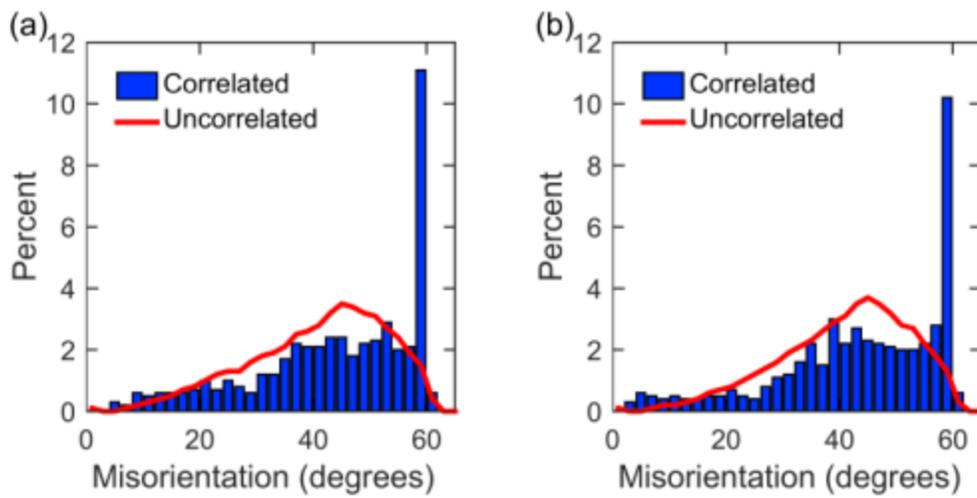

**Figure 8:** Misorientation angle distributions for (a) sputtered and (b) electrodeposited Ni, both of ~20 nm grain size. The correlated data set measures the misorientation along real boundaries, and the uncorrelated is between random grains.

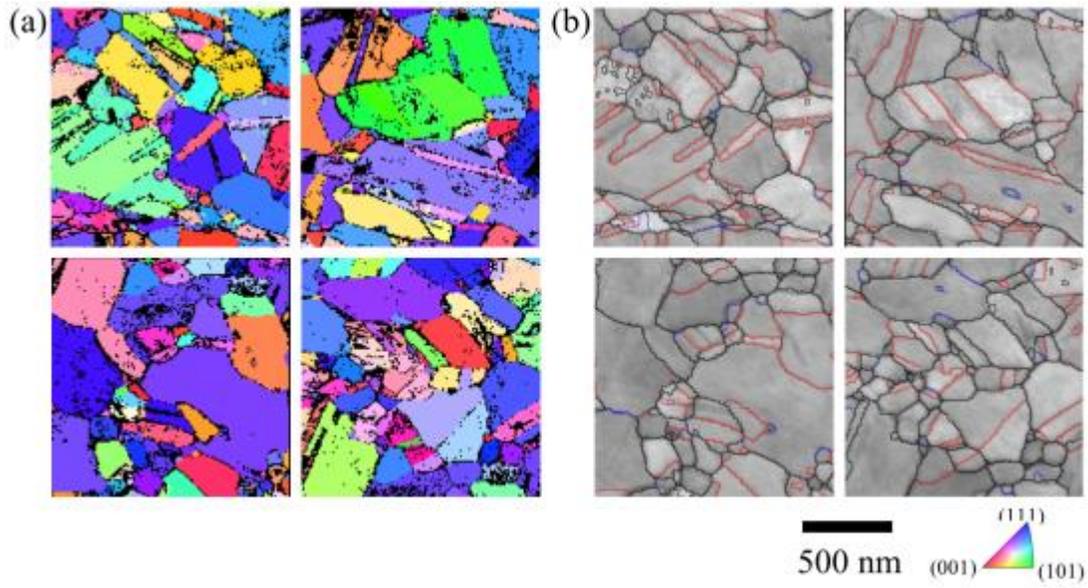

**Figure 9: (a) Grain orientation maps and (b) reconstructed grain boundaries for the annealed sputtered material. Red lines are Σ3 boundaries, blue are Σ1s and all other boundaries are shown in black. The grey scale contrast is the Kikuchi pattern quality.**

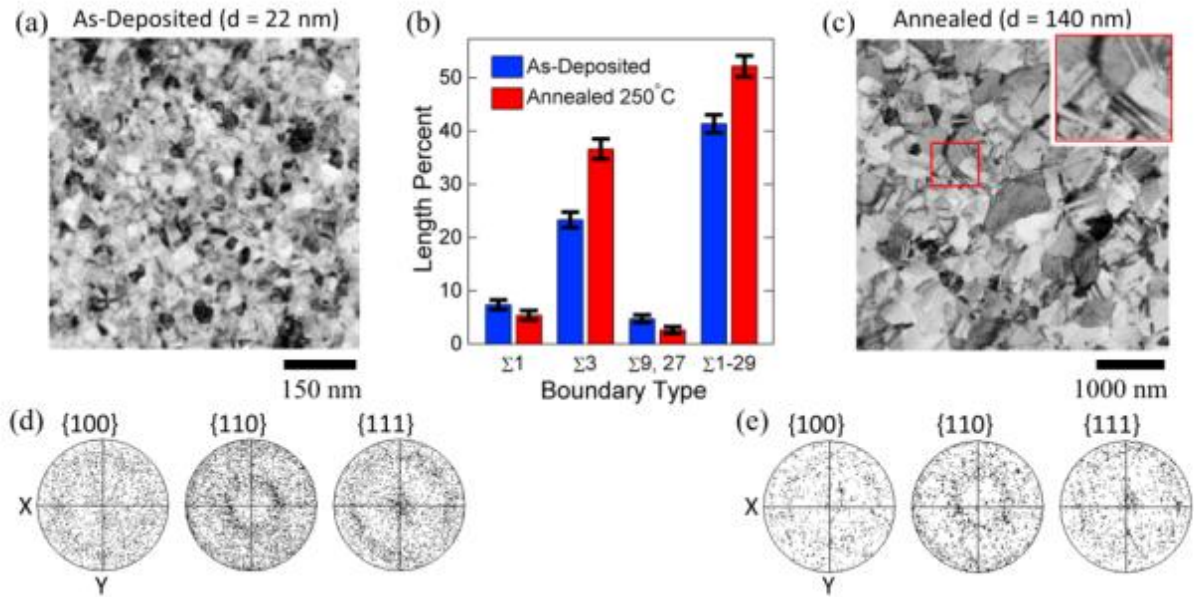

**Figure 10: Bright field TEM images and TKD pole figures for sputtered samples having (a, d) 22 nm and (c, e) 140 nm grain sizes along with their GBCDs (b). The 140 nm grain size sample in parts (c) and (e) was annealed from the as-deposited state shown in (a) and (d). Inset in (c) shows multiple growth twins.**

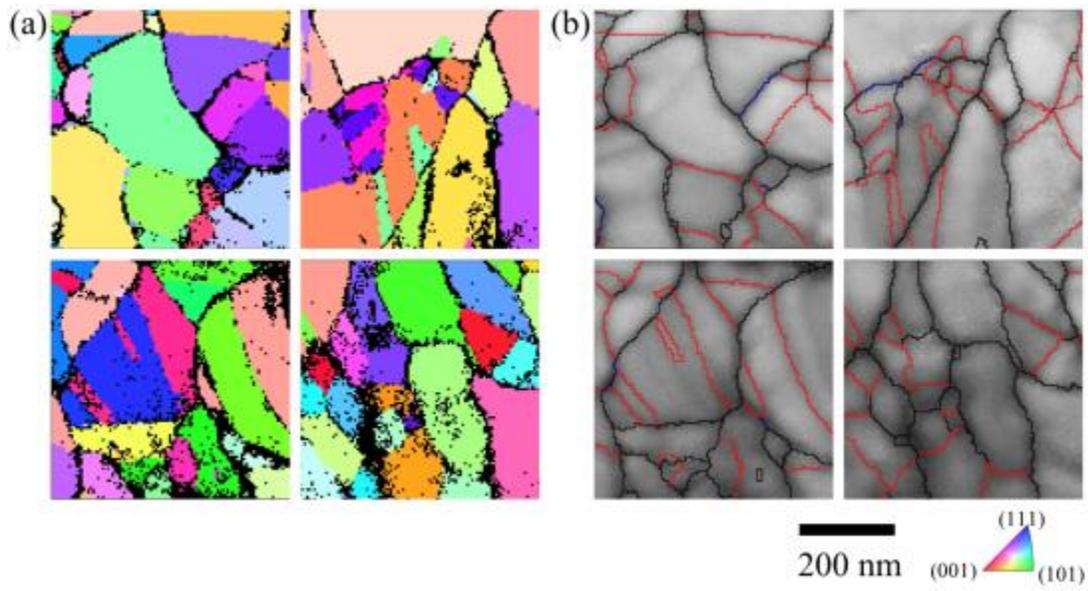

**Figure 11: (a) Grain orientation maps and (b) reconstructed grain boundaries for the annealed sputtered material. Red lines are Σ3 boundaries, blue are Σ1s and all other boundaries are shown in black. The grey scale contrast is the Kikuchi pattern quality.**

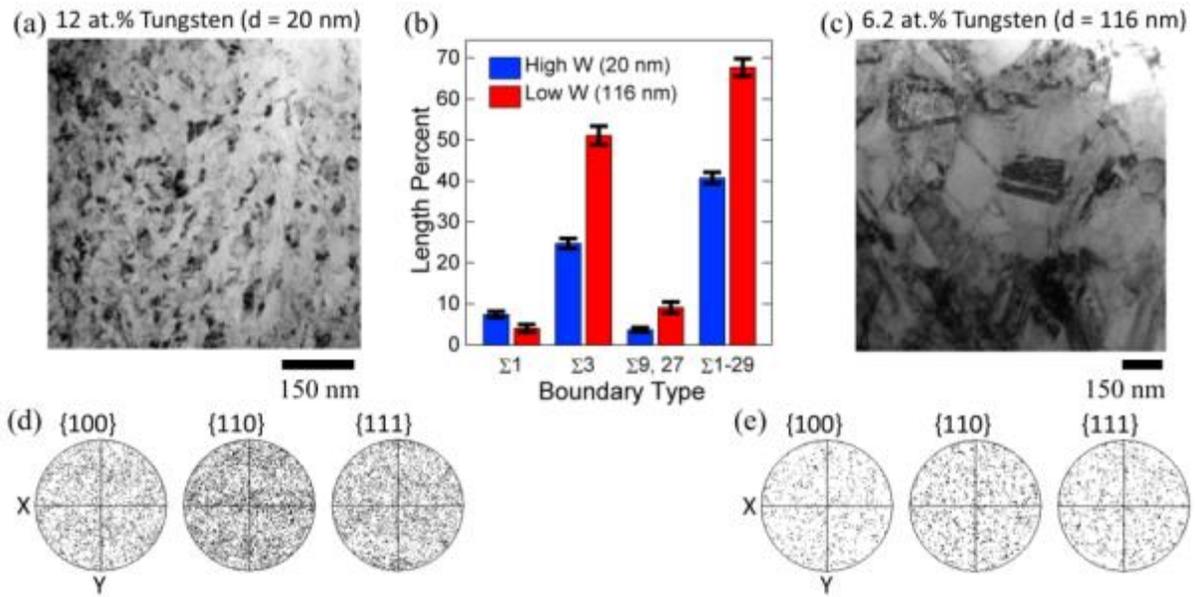

**Figure 12: Bright field TEM images and TKD pole figures for electrodeposited samples with (a, d) 20 nm and (c, e) 115 nm grain sizes, along with their GBCDs (b). The grain size was controlled via the W content using reverse pulse plating.**

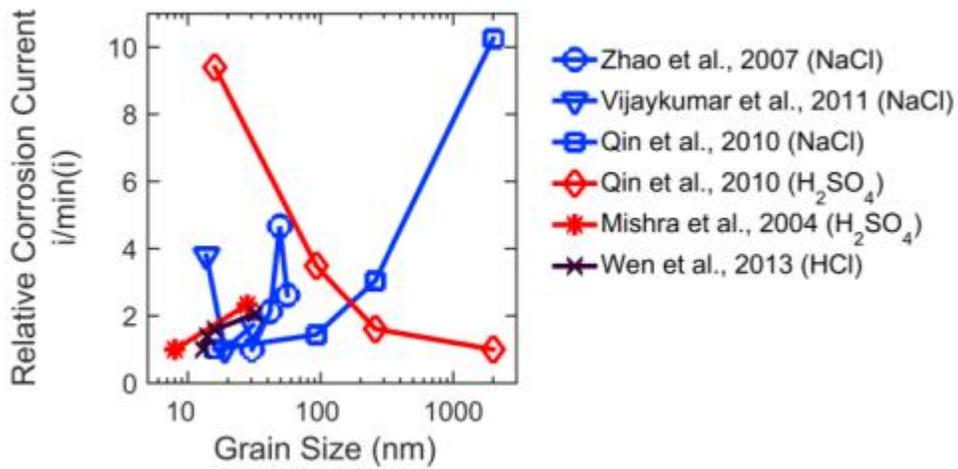

**Figure 13: The normalized corrosion currents reported for electrodeposited Ni in several electrolytes and spanning the nanocrystalline and ultrafine grained regimes. The data for NaCl (blue lines) and $H_2SO_4$ (red lines) both show self-conflicting trends. All data shown is from prior literature.[76-80]**

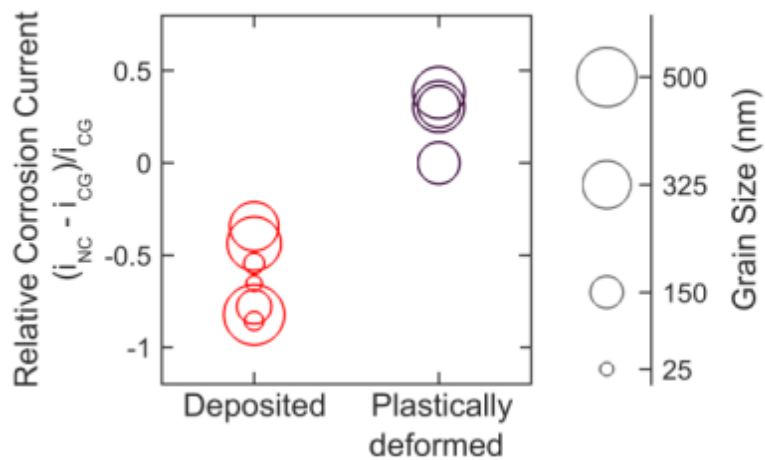

**Figure 14: The effect of processing method on the corrosion of nanocrystalline copper manufactured by either deposition or severe plastic deformation. The fractional difference between the nanocrystalline and coarse grained corrosion currents from each study is shown, with negative ordinate values indicating an improvement in corrosion resistance. The area of each marker is proportional to that material's grain size. All data shown is from prior literature.[75, 81-85]**